\documentclass[twocolumn,prc,amssymb, aps,superscriptaddress, 
amsmath]{revtex4}

\usepackage{color}
\usepackage{amsmath,mathtools,booktabs,
            microtype} 

\usepackage{graphicx}
\usepackage{dcolumn}
\usepackage{bm}
\usepackage{multirow}
\usepackage{times}
\usepackage{url}
\usepackage{tikz}
\usepackage[version=4]{mhchem}

\usetikzlibrary{arrows,decorations.pathmorphing}
\usetikzlibrary{arrows.meta}
\usetikzlibrary{calc}

\definecolor{violet}{HTML}{602969}
\definecolor{red}{HTML}{FC0009}
\definecolor{orange}{HTML}{FF6319}
\definecolor{green}{HTML}{00933C}
\definecolor{blue}{HTML}{0036A6}
\definecolor{yellow}{HTML}{FFBE00}
\definecolor{lightgrey}{HTML}{A7A9AC}

\newcommand{\be}{\begin{equation}}
\newcommand{\ee}{  \end{equation}}
\newcommand{\ba}{\begin{eqnarray}}
\newcommand{\ea}{  \end{eqnarray}}

\begin{document}

\title{Out-of-Time Ordered Correlator for a Chaotic Many-Body Quantum
  System}

\author{Hans A. \surname{Weidenm\"uller}}
\email{haw@mpi-hd.mpg.de}
\affiliation{Max-Planck-Institut f\"ur Kernphysik, Saupfercheckweg 1,
  D-69117 Heidelberg, Germany}

\date{\today}


\begin{abstract}Using the parametric representation of a chaotic many-body
  quantum system derived earlier, we calculate explicitly the
  large-time dependence and asymptotic value of the out-of-time
  correlator (OTOC) of that system. The dependence on time $t$ is
  determined by $\Delta t / \hbar$. Here $\Delta$ is the energy
  correlation width within which the Bohigas-Giannoni-Schmit
  conjecture applies. We conjecture that $\Delta$ is universally
  related to the leading Ljapunov coefficient of the corresponding
  classical system by $\Delta = \hbar \lambda_{\max}$. Then the
  large-time behavior of OTOC is given by the dimensionless parameter
  $\lambda_{\max} t$.
\end{abstract}

\maketitle


\section{Introduction}
\label{int}

After the early work of Larkin and Ovchinnikov~\cite{Lar69},
out-of-time ordered correlators (OTOC) have widely been used in recent
years to investigate and characterize chaos in quantum systems. A
thorough review has recently been given in Ref.~\cite{Gar23}. The
interest in OTOC was initially triggered by the hope that for short
times, the rate of growth with time of OTOC could be indicative of
quantum chaos. It turned out, however, that this expectation does not
hold universally, and different time regimes have been studied
since~\cite{Gar23}.

In the present paper we calculate the out-of-time ordered correlator
of a chaotic many-body quantum system, using the universal
parametrization of the Hamiltonian of such a system established in
Ref.~\cite{Wei24}. We adopt the notation of Ref.~\cite{Mal16}. We show
that for large times, the time dependence of the correlator is
determined by a parameter that characterizes the chaotic quantum
system. We conjecture that this parameter, in turn, is given by the
leading Ljapunov coefficient of the corresponding classical system.

\section{Definitions}
\label{def}

Given two simple operators $V$ and $W$ as defined in
Ref.~\cite{Mal16}, the out-of-time ordered correlator $C(t)$ of a
system governed by the Hamiltonian $H$ is defined as the thermal
average of the square of the commutator of $V$ and $W(t)$,
\ba
\label{d1}
C(t) = - \frac{1}{{\rm Tr} (Z)} {\rm Tr} \big( Z [W(t), V]^2 \big) \ .
\ea
Here $Z$ is the canonical partition function,
\ba
\label{d2}
Z = \exp \{ - \beta H \} \ ,
\ea
and the time dependence of $W(t)$ is given by
\ba
\label{d3}
W(t) = \exp \{ - i H t \} W \exp \{ i H t \} \ .
\ea
We use $\hbar = 1$ throughout, put
\ba
\label{d4}
U(t) = \exp \{ - i t H \} \ ,
\ea
and write Eq.~(\ref{d1}) in the form
\ba
\label{d5}
C(t) &=& - \frac{1}{{\rm Tr} (Z)} {\rm Tr} \bigg[ Z \bigg( V U(t) W
  U^\dag(t) V U(t) W U^\dag (t) \nonumber \\
  && + U(t) W U^\dag(t) V  U(t) W U^\dag(t) V \nonumber \\
  && - W U^\dag(t) V^2 U(t) W - V U(t) W^2 U^\dag(t) V \bigg) \bigg] \ .
\ea
We have used that $U$ commutes with $Z$. Following Ref.~\cite{Mal16},
we consider in addition to $C(t)$ the symmetrized function
\ba
\label{d6}
{\cal F}(t) = \frac{1}{{\rm Tr} (Z)} {\rm Tr} \bigg( V Y(\chi) W
Y^\dag(\chi) V Y(\chi) W Y^\dag(\chi) \bigg) \ .
\ea
The matrix $Y(\chi)$ is defined as
\ba
\label{d8}
Y(\chi) = \exp \{ \chi H \} \ .
\ea
In Eq.~(\ref{d6}), $\chi$ is given by
\ba
\label{d7}
\chi = - \beta / 4 - i t \ .
\ea
With that choice of $\chi$, the matrix $Y(\chi)$ is not unitary but
obeys
\ba
\label{d9}
Y(\chi) Y^\dag(\chi) = \exp \{ - \beta H / 2 \} \ ,
\ea
in contrast to the unitary matrix $U(t)$ defined in
Eq.~(\ref{d4}). The partition function is given by
\ba
\label{d10}
Z = Y(- \beta) \ .
\ea
While the first (second) term on the right-hand side of Eq.~(\ref{d5})
carries the full partition function $Z$ sandwiched between adjacent
factors $(W, V)$ ($(V, W)$, respectively), the function ${\cal F}(t)$
carries the fourth root $Z^{1/4}$ of $Z$ sandwiched between every
adjacent pair of factors $(V, W)$ and $(W, V)$. As done in
Ref.~\cite{Mal16}, we consider ${\cal F}(t)$ as paradigmatic for the
expressions ocurring in Eq.~(\ref{d5}). Use of the symmetrized
form~(\ref{d6}) reduces the number of terms and simplifies the
bookkeeping of the calculation. We study both $C(t)$ and ${\cal
  F}(t)$.

\section{Chaotic Many-Body Quantum System}
\label{cha}

For the representation of the Hamiltonian $H$ of the chaotic many-body
quantum system we follow Ref.~\cite{Wei24}. We assume that the system
is invariant under time reversal. We write $H$ as the sum of an
integrable Hartree-Fock term $H_{\rm HF}$ and a residual interaction
$V$,
\ba
\label{h1}
H = H_{\rm HF} + V \ .
\ea
The eigenstates (eigenvalues) of $H_{\rm HF}$ are labeled $(m, n,
\ldots)$ ((${\cal E}_m, {\cal E}_n, \ldots$), respectively). The
interaction $V$ mixes the eigenstates of $H_{\rm HF}$. In the
Hartree-Fock basis, diagonalization of $H$ yields
\ba
\label{h2}
H_{m n} = \sum_\alpha O_{m \alpha} E_\alpha O_{n \alpha} \ .
\ea
If the system is chaotic, the Bohigas-Giannoni-Schmit (BGS)
conjecture~\cite{Boh84} applies. The conjecture says that the
eigenvalues $E_\alpha$ and eigenvectors $O_{m \alpha}$ of a chaotic
quantum system locally obey GOE statistics. Here GOE stands for the
Gaussian Orthogonal Ensemble of random matrices~\cite{Meh04}. The term
``local'' means that GOE statistics holds in a finite interval of
width $\Delta$ only. Eigenvalues and eigenfunctions separated by an
energy interval bigger than $\Delta$ are statistically
uncorrelated. We recall that in the framework of random-matrix
theory~\cite{Meh04}, GOE statistics is derived in the limit of
infinite GOE matrix dimension. Therefore, quantitative agreement with
GOE predictions can be expected only in the semiclassical regime of
high excitation energy $E$ and of sufficiently high level density
$\rho(E)$ where the number
\ba
\label{h2a}
N = \Delta \rho(E)
\ea
of eigenvalues within $\Delta$ obeys $N \gg 1$. In that regime, the
elements $O_{m \alpha}$ of the diagonalizing orthogonal matrix possess
a zero-centered Gaussian distribution, and the eigenvalues $E_\alpha$
obey Wigner-Dyson statistics. Eigenvalues and eigenfunctions are
statistically independent.

To make these statistical properties analytically tractable we follow
Ref.~\cite{Wei24} further. Following standard practice in statistical
mechanics, we consider instead of the actual Hamiltonian $H$ in
Eq.~(\ref{h2}) an ensemble of Hamiltonians. The use of an ensemble has
the advantange that averages are taken over the ensemble (and not as
running averages over the spectral distribution of eigenvalues and
matrix elements of a fixed Hamiltonian). Without change of notation,
the ensemble is defined by promoting the ordinary variables $O_{m
  \alpha}$ and $E_\alpha$ in Eq.~(\ref{h2}) to random variables. The
matrix elements $O_{m \alpha}$ become zero-centered Gaussian random
variables with second moments
\ba
\label{h3}
\langle O_{m \alpha} O_{n \beta} \rangle = \delta_{m n} \delta_{\alpha
    \beta} F({\cal E}_m - \overline E_\alpha) \ .
\ea
The angular brackets denote the average over the ensemble. The
function $F$ is a normalized Gaussian of width $\Delta$,
\ba
\label{h4}
&& F({\cal E}_m - \overline E_\alpha) \nonumber \\
&& = \frac{1}{\sqrt{2 \pi} \rho({\cal E}_m) \Delta}
\exp \{ - ({\cal E}_m - \overline{E}_\alpha)^2 / ( 2 \Delta^2) \} \ .
\ea
Here $\overline{E_\alpha}$ is the average of $E_\alpha$ over the
ensemble. Within the correlation width $\Delta$, the eigenvalues
$E_\alpha$ obey Wigner-Dyson statistics. The random variables $O_{m
  \alpha}$ and $E_\alpha$ are statistically independent. In other
words, the matrix elements $O_{m \alpha}$ and eigenvalues $E_\alpha$
locally follow GOE statistics~\cite{Meh04}. These postulates define
the random-matrix ensemble $H$ via Eq.~(\ref{h2}). We use that
ensemble in the definitions in Section~\ref{def}, too. Then the
functions $C(t)$ and ${\cal F}(t)$ become time-dependent stochastic
processes.

In Ref.~\cite{Wei24}, it is emphasized that the
representation~(\ref{h2}) of $H$ in terms of random variables is valid
only after a characteristic time that is needed for transients to die
out and for the statistical properties to become dynamically
observable. That is why the present approach cannot deal with the
short times that were in the focus of earlier investigations, see
Ref.~\cite{Gar23}.

\section{Approach}

For the stochastic processes $C(t)$ and ${\cal F}(t)$ defined in
Eqs.~(\ref{d5}) and (\ref{d6}), we calculate both the ensemble average
and the variance in the HF basis. As in Ref.~\cite{Wei24} we assume
that the mixing of HF states due to $V$ is a local phenomenon, so that
the average level densities of the HF Hamiltonian and of the full
Hamiltonian $H$ in Eq.~(\ref{h1}) are the same. We use the symbol
$\rho(E)$ indiscriminately for both.

In the ensemble average of ${\cal F}(t)$, written as
\ba
\label{a1}
\bigg\langle \frac{1}{{\rm Tr} (Z)} {{\rm Tr} \bigg( V Y(\chi) W
  Y^\dag(\chi) V Y(\chi) W Y^\dag(\chi)
\bigg)} \bigg\rangle \ ,
\ea
the stochastic element is the operator $Y(\chi)$ defined in
Eq.~(\ref{d8}), with $\chi$ given by Eq.~(\ref{d7}) for the terms in
the numerator of expression~(\ref{a1}) and by $\chi = - \beta$ for $Z$
in the denominator. In the HF basis, the matrix elements of the
operator $Y(\chi)$ are
\ba
\label{a2}
Y_{m n}(\chi) &=& \sum_\alpha O_{m \alpha} \exp \{ \chi E_\alpha \}
O_{n \alpha} \ .
\ea
Analogously and with proper choices of the arguments $\chi$, the
elements of $C(t)$ in Eq.~(\ref{d5}) can be written in terms of the
operators $Y(\chi)$ and $Y^\dag(\chi)$. The average in
expression~(\ref{a1}) has to be carried out independently over the
zero-centered Gaussian-distributed matrix elements $O_{m \alpha}$ and
over the eigenvalues $E_\alpha$ that follow a Wigner-Dyson
distribution. We work in the semiclassical regime of high level
density where the number $N$ of states in the interval $\Delta$
defined in Eq.~(\ref{h2a}) obeys $N \gg 1$. We keep only terms of
leading order in $1 / N$.

\section{Moments}
\label{mom}

Prior to evaluating $\langle C(t)$, $\langle {\cal F}(t) \rangle$ and
the variances, we calculate in this Section moments of the matrix
elements of $Y(\chi)$ in Eq.~(\ref{a2}). We use that $Y^\dag(\chi) =
Y(\chi^*)$. The moment of order $k$ is the ensemble average of the
product
\ba
\label{m1}
\prod_{j = 1}^k Y_{m_j n_j} (\chi_j) = \prod_{j = 1}^k \sum_{\alpha_j}
O_{m_j \alpha_j} \exp \{ \chi_j E_{\alpha_j} \} O_{n_j \alpha_j} \ .
\ea
In Eq.~(\ref{m1}) we allow for different arguments $\chi_j$ of
$Y$. That is done to account for the fact that the matrices $Y$ appear
with different arguments in both Eqs.~(\ref{d5}) and
(\ref{d6}). Formally, the $k$ factors in expression~(\ref{m1})
commute. However, every such factor stands for a factor that actually
occurs in the expression~(\ref{a1}) for $\langle {\cal F}(t) \rangle$
or the corresponding expression for $\langle C(t) \rangle$. Interlaced
between these factors are the matrices $W_{m n}$ or $V_{m n}$, as the
case may be. In calculating the average of expression~(\ref{m1}) we
must, therefore, respect the order of the factors labeled $j = 1,
\ldots, k$.

The average of expression~(\ref{m1}) is calculated by averaging first
over the elements $O_{m \alpha}$ of the orthogonal matrices. These are
zero-centered Gaussian-distributed random variables. The average over
a product of such variables is obtained by averaging all possible
pairs using Eqs.~(\ref{h3}) and (\ref{h4}). The average of a pair is
called a contraction and is pictorially indicated by an overbar (a
``contraction line'') that starts with the first and ends with the
second factor comprising the pair. The contraction lines connecting
all pairs of matrix elements $O_{m \alpha}$ in a product of the
form~(\ref{m1}) constitute a contraction pattern. In Ref.~\cite{Mon75}
it is shown that to leading order in $1 / N$, the only allowed
contraction patterns are the ones where no two contraction lines
intersect. For instance, in the average over the matrices $O$ of ${\rm
  Tr} ( O A O B O C O D)$ where $A, B, C, D$ are operators that are
independent of the matrices $O$, the contraction pattern where the
first factor $O$ is contracted with the third one and the second
factor $O$ with the fourth one, is excluded.

It is convenient to calculate first the correlated part of the
moment~(\ref{m1}). That part is defined as containing only contraction
patterns where each factor $Y$ in the product~(\ref{m1}) can be
reached from every other such factor by following a sequence of
contraction lines within the pattern. In other words, the correlated
part of the moment does not consist of two or more parts that are not
connected to each other by contraction lines. Inspection shows that
for the average of the product in expression~(\ref{m1}), only a single
contraction pattern contributes to the correlated part. That is the
pattern where $O_{n_1 \alpha_1}$ is contracted with $O_{m_2
  \alpha_2}$, $O_{n_2 \alpha_2}$ with $O_{m_3 \alpha_3}$, $\ldots$,
$O_{n_k \alpha_k}$ with $O_{m_1 \alpha_1}$. For the
product~(\ref{m1}), that fact drastically reduces the potentially
large number of contraction patterns to a single one and makes it
possible to write down the totally correlated part of
expression~(\ref{m1}).

Our definition of the correlated part of the product~(\ref{m1}) refers
to the stochastic variables $O_{m \alpha}$ only. Correlations may be
due, however, also to the stochastic variables $E_\alpha$. For
instance, two parts of expression~(\ref{m1}), each intrinsically
correlated according to our definition, may not be connected by
contraction lines and, in the sense of our definition, be not
correlated with one another. Then averaging over the stochastic energy
variables $E_\alpha$ and $E_\beta$ occuring in either of the two parts
will establish a correlation between the two parts owing to the
correlations between $E_\alpha$ and $E_\beta$. In the GOE, the
two-point correlation function of the energy variables $E_\alpha$ and
$E_\beta$, a function of $E_\alpha - E_\beta$, is significantly
different from zero only over an energy range given by low multiples
of the mean level spacing $d$. For $N \gg 1$, that range is negligible
compared to $N d$, see Refs.~\cite{Meh04, Wei24}.  Therefore,
correlations of two parts of the product~(\ref{m1}) induced only by
energy correlations are small of order $1 / N$ compared to the product
of the uncorrelated parts and, therefore, negligible. It follows that
our definition of correlations fully applies also when energy
correlations are taken into account. Moreover, after averaging over
the stochastic variables $O_{m \alpha}$ we may, to leading order in $1
/ N$, replace the energies $E_\alpha$ by their mean values
$\overline{E_\alpha}$. The remaining sums over the eigenvalues
$\overline{E_\alpha}$ can then be done as in Ref~\cite{Wei24}.

The first moment of $Y_{m n}(\chi)$ is
\ba
\label{m2}
\langle Y_{m n}(\chi) \rangle = \delta_{m n} \exp \{ \chi^2 \Delta^2 / 2
+ \chi {\cal E}_m \} \ .
\ea
More generally, the totally correlated part of expression~(\ref{m1})
is, for positive integer $k$, given by
\ba
\label{m3}
&& \langle \prod_{j = 1}^k Y_{m_j n_j}(\chi_j) \rangle_{\rm corr} =
\frac{\sqrt{1/k}}{(\sqrt{2 \pi} \rho({\cal E}_{m_1}) \Delta)^{k - 1}}
\nonumber \\
&& \times \bigg( \prod_{j = 1}^k \delta_{n_j m_{j + 1}} \bigg) \exp \{ \chi
\sum_{j = 1}^k {\cal E}_{m_j} / k + \chi^2 \Delta^2 / ( 2 k) \}
\nonumber \\
&& \times \exp \{ \sum_{j = 1}^{k - 1} \sum_{l > j}^k ({\cal E}_{m_j} -
{\cal E}_{m_l})^2 / ( 2 k \Delta^2) \} \ .
\ea
We have put
\ba
\label{m4}
\chi = \sum_{j = 1}^k \chi_j \ .
\ea
Here and in what follows, the index ${\rm corr}$ denotes the totally
correlated part.

To justify Eq.~(\ref{m3}), we note that on the left-hand side of
Eq.~(\ref{m3}) there occur $(2 k)$ factors $O_{m \alpha}$, see
Eq.~(\ref{a2}). According to Eqs.~(\ref{h3}) and (\ref{h4}) and to the
rules stated above, contraction gives rise on the right hand side to
the product of $k$ Kronecker deltas referring to HF states, to $k$
factors $1 / (\sqrt{2 \pi} \rho({\cal E}_m) \Delta)$, to $k$ factors
$F$, and to $k$ Kronecker deltas referring to states $\alpha$. The
latter force the energy argument $E_\alpha$ to be the same in all
functions $F$. The Gaussian integral over $E_\alpha$ removes one
factor $1 / (\sqrt{2 \pi} \rho({\cal E}_m) \Delta)$ and yields the two
exponential factors and the factor $1 / \sqrt{k}$.

\section{Average Partition Function}
\label{par}

In view of the appearance of the partition function $Z$ in the
denominators of Eqs.~(\ref{d5}) and (\ref{d6}), the calculation of
$\langle C(t) \rangle$ and of $\langle {\cal F}(t) \rangle$ looks like
a challenging task. We now show that for $N \gg 1$, ${\rm Tr} (Z)$ may
be replaced by its average value $\langle {\rm Tr} (Z) \rangle$.

From Eq.~(\ref{m2}) with $\chi = - \beta$ we have
\ba
\label{p1}
\langle {\rm Tr} (Z) \rangle = \exp \{ \beta^2 \Delta^2 / 2 \}
{\rm Tr} (Z_{\rm HF}) \ .
\ea
Here
\ba
\label{p2}
Z_{\rm HF} = \sum_m \exp \{ - \beta {\cal E}_m \}
\ea
is the partition function of the Hartree-Fock Hamiltonian. There are
two contributions to the correlated part of $[{\rm Tr} (Z)]^2$. One
arises from the correlations of the matrices $O_{m \alpha}$. With
$\chi_1 = - \beta = \chi_2$, Eq.~(\ref{m3}) yields for $k = 2$
\ba
\label{p3}
\langle [{\rm Tr} (Z)]^2 \rangle_{\rm corr} &=& \exp \{ \beta^2 \Delta^2
\} \nonumber \\
&& \times \sum_m \frac{1}{2 \sqrt{\pi} \rho({\cal E}_m) \Delta} \exp \{
- 2 \beta {\cal E}_m \} \ .
\ea
That contribution is small of order $1 / N^2$ compared to $\langle
{\rm Tr} (Z) \rangle^2$. The second contribution arises from the
correlation of eigenvalues $E_\alpha$ and $E_\beta$ in the two factors
${\rm Tr} (Z)$. It follows from the statements in the paragraph
preceding Eq.~(\ref{m2}) that that contribution is small of order $1 /
N$ compared to $\langle {\rm Tr} (Z) \rangle^2$. In the large $N$
limit, we may, therefore, replace in expression~(\ref{a1}) and in the
corresponding expression for $\langle C(t) \rangle$ the factor $[{\rm
    Tr} (Z)]^{- 1}$ by $\langle {\rm Tr} (Z) \rangle^{- 1}$ with
$\langle {\rm Tr} (Z) \rangle$ given by Eq.~(\ref{p1}). The remaining
averages of $C(t)$ and ${\cal F}(t)$ are then taken only over the
traces in the numerators. The same statement applies to the variances.

It may be disconcerting that $\langle {\rm Tr} (Z) \rangle$ differs
from ${\rm Tr} (Z_{\rm HF})$ by the factor $\exp \{ \beta^2 \Delta^2 /
2 \}$. The factor results from the Gaussian used for $F({\cal E}_m -
\overline{E}_\alpha)$ in Eq.~(\ref{h4}). The normalization of $F({\cal
  E}_m - \overline{E}_\alpha)$ guarantees the correct normalization of
the orthogonal matrices $O_{m \alpha}$. It cannot guarantee, however,
the expected normalization of $\langle {\rm Tr} (Z) \rangle$. It turns
out that the factor $\exp \{ \beta^2 \Delta^2 / 2 \}$ cancels
completely in the expression for $\langle C(t) \rangle$ and at least
partially in the expression for $\langle {\cal F}(t) \rangle$, see
Eqs.~(\ref{av4}) and (\ref{av5}).

\section{Averages}
\label{ave}

With ${\rm Tr} (Z)$ replaced by $\langle {\rm Tr} (Z) \rangle$, it
remains to calculate the averages of the terms in the numerators of
Eqs.~(\ref{d5}) and (\ref{d6}). We illustrate the way averages are
taken by using the numerator of Eq.~(\ref{d6}) as example. The average
is written as the sum over products of correlated moments. The
following possibilities exist. (i) Each of the four factors $Y$ is
replaced by its average (one possibility); (ii) two factors $Y$ are
replaced by the correlated moment, the other two by their mean
values (six possibilities); (iii) two factors $Y$ are replaced by
the correlated moment and likewise for the remaining two factors $Y$
(two possibilities); (iv) three factors $Y$ are replaced by the
correlated third moment, the fourth factor by its average (four
possibilities); (iv) the four factors $Y$ are replaced by the
correlated fourth moment (one possibility). That gives 14 terms for
Eq.~(\ref{d6}) and a substantially larger number for Eq.~(\ref{d5}).

The resulting expressions drastically simplify if we follow the
example of Ref.~\cite{Mal16}. There, it is assumed that $V$ and $W$
have zero one-point functions. In the present context that assumption
reads
\ba
\label{av1}
&& \frac{1}{Z} \sum_m V_{m m} \exp \{ - \beta {\cal E}_m \} = 0 \ ,
\nonumber \\
&&\frac{1}{Z} \sum_m W_{m m} \exp \{ - \beta {\cal E}_m \} = 0 \ .
\ea
Here we use a slightly more stringent assumption by postulating that
for $A= V$ and $A = W$, fixed ${\cal E}_n$, any positive $\beta$, and
arbitrary positive integer values of $k$ we have
\ba
\label{av2}
\sum_m A_{m m} \exp \{ - \beta {\cal E}_m \} \exp \{ - ({\cal E}_m -
{\cal E}_n)^2 / (2 k \Delta^2) \} = 0 \ . \nonumber \\
\ea
In comparison with Eq.~(\ref{av1}), Eq.~(\ref{av2}) postulates that
the thermal average of $A$ vanishes already when taken only over an
energy interval of width $\Delta$.

Terms of the form~(\ref{av2}) arise in averaging the numerator terms
in $C(t)$ and in ${\cal F}(t)$. Eq.~(\ref{m3}) shows that such is the
case whenever the term contains a product of the form $Y^\dag V Y$ or
$Y W Y^\dag$ wherein both factors $Y$ either form a correlated second
moment or are both members of the same correlated moment of higher
order. According to Eq.~(\ref{av2}), all these terms vanish.

For the numerators of the first two terms of Eq.~(\ref{d5}) and for
the numerator of Eq.~(\ref{d6}) that leaves us with possibility~(i)
and possibility~(ii) as the only remaining options. There are two
contributions from possibility~(ii). These arise from replacing two
factors $Y$ by their correlated moment and the remaining two factors
$Y^\dag$ by their individual averages, and conversely. That gives a
total of three terms. The three terms can be worked out with the
help of Eq.~(\ref{m3}). It turns out that for $\langle C(t) \rangle$,
each of these terms carries the factor $\exp \{ - 2 t^2 \Delta^2 \}$
while for $\langle {\cal F}(t) \rangle$, each of these terms carries
the factor $\exp \{ - 2 t^2 \Delta^2 - 3 \beta^2 \Delta^2 / 8 \}$. For
clarity and to save space we write the totality of these terms for
$\langle C(t) \rangle$ as $\exp \{ - 2 t^2 \Delta^2 \} C_0$ and for
$\langle {\cal F}(t) \rangle$ as $\exp \{ - 2 t^2 \Delta^2 - 3 \beta^2
\Delta^2 / 8 \} {\cal F}_0$. The explicit expression for ${\cal F}_0$
is given in the Appendix. Except for a rearrangement of factors $Z$,
an analogous expression holds for $C_0$ and is not given.  For the two
terms in the last line of Eq.~(\ref{d5}), the only remaining
possibilities are $Y \to \langle Y \rangle$ for every factor $Y$ or
$Y^\dag$, and $Z \to \langle Z \rangle, U^\dag V^2 U \to (U^\dag V^2
U)_{\rm corr}, U W^2 U^\dag \to (U W^2 U^\dag)_{\rm corr}$. These
terms, too, are readily worked out with the help of Eq.~(\ref{m3}).

We find
\ba
\label{av4}
&&\langle C(t) \rangle = - \exp \{ - 2 t^2 \Delta^2 \} \frac{1}
{{\rm Tr} (Z_{\rm HF})}C_0 \nonumber \\
&& + \frac{1}{{\rm Tr} (Z_{\rm HF})} \bigg[ \sum_m (W Z_{\rm HF} W)_{m m}
\bigg( \sum_l \frac{1}{\sqrt{2 \pi} \rho({\cal E}_m)  \Delta}
\nonumber \\
&& \qquad \qquad \times \exp \{ - ({\cal E}_m - {\cal E}_l)^2 / 4 \Delta^2
\} (V^2)_{l l} \bigg) \nonumber \\
&& \qquad + \sum_m (V Z_{\rm HF} V)_{m m} \bigg( \sum_l \frac{1}
{\sqrt{2 \pi} \rho({\cal E}_m)  \Delta} \nonumber \\
&& \qquad \qquad \times \exp \{ - ({\cal E}_m - {\cal E}_l)^2 / 4 \Delta^2
\} (W^2)_{l l} \bigg) \bigg]
\ea
and
\ba
\label{av5}
&& \langle {\cal F}(t) \rangle = \exp \{ - 2 t^2 \Delta^2 - 3 \beta^2
\Delta^2 / 8\} \frac{1}{{\rm Tr} (Z_{\rm HF})} {\cal F}_0 \ .
\ea
The decay in time of both $\langle C(t) \rangle$ and $\langle {\cal
  F}(t) \rangle$ is governed by a Gaussian of width $(2 \Delta)^{-
  1}$. Both $C_0$ and ${\cal F}_0$ contain terms that oscillate in
time.

\section{Fluctuations}
\label{var}

To show that the fluctuations of ${\cal F}(t)$ and of $C(t)$ around
their averages are negligible, we show that the variances of ${\cal
  F}(t)$ and of $C(t)$ are small of order $1 / N$ compared to the
squares of the mean values of these expressions. In Section~\ref{par}
we have shown that for $N \gg 1$ we may replace ${\rm Tr} (Z)$ by
$\langle {\rm Tr} (Z) \rangle$. Therefore, we confine our proof to the
numerators on the right-hand sides of Eqs.~(\ref{d5}) and (\ref{d6}).

We begin with ${\cal F}(t)$. The variance of the numerator in
Eq.~(\ref{d6}) is given by
\ba
\label{v1}
&& \bigg\langle {\rm Tr} \bigg( V Y(\chi) W Y^\dag(\chi) V Y(\chi) W
Y^\dag(\chi) \bigg) \nonumber \\
&& \ \ \ \times {\rm Tr} \bigg( V Y(\chi) W Y^\dag(\chi) V Y(\chi) W
Y^\dag(\chi) \bigg) \bigg\rangle_{\rm corr} \ .
\ea
We wish to show that expression~(\ref{v1}) is small of order $1 / N$
compared to the product $\Pi$ of the averages of the two
traces. Expression~(\ref{v1}) differs from zero if at least two
eigenvalues $({\cal E}_m, {\cal E}_n)$ ocurring in either trace are
correlated, or if at least one pair of matrix elements $(O_{m \alpha},
O_{n \beta})$ located in either trace is contracted. In the first
case, it suffices to recall that in the text preceding Eq.~(\ref{m2})
it is shown that the correlation of two eigenvalues is small of order
$N^{- 1}$. In the second case, we express both the correlated
expression~(\ref{v1}) and each factor in the product $\Pi$ of the
averages of the two traces as a sum of traces over products of totally
correlated moments. To every term in the resulting sum in $\Pi$ there
belongs at least one term in expression~(\ref{v1}). In that term, at
least one correlated moment $M_1$ of order $k_1$ in the first trace
and one correlated moment $M_2$ of order $k_2$ in the second trace in
$\Pi$ have been contracted to yield a correlated moment $M_3$ of order
$k_3 = k_1 + k_2$ in expression~(\ref{v1}). Conversely, each term in
expression~(\ref{v1}) must contain at least one correlated moment
$M_3$ with contributions from either trace. That moment may be viewed
as due to the contraction of two moments $M_1$ and $M_2$ in either
trace. To show that in every such case, the contribution of $M_3$ is
small of order $1 / N$ compared to the combined contributions of $M_1$
and $M_2$, we use Eq.~(\ref{m3}).  The equation shows that in each
correlated moment of order $k$ there occur $k$ Kronecker deltas and
$(k - 1)$ factors $1 / N$. The number $k_1$ of Kronecker deltas in
$M_1$ plus the number $k_2$ of Kronecker deltas in $M_2$ is equal to
the number $k_3 = k_1 + k_2$ of Kronecker deltas in $M_3$. But the
number $(k_1 - 1)$ of factors $1 / N$ in $M_1$ plus the number $(k_2 -
1)$ of such factors in $M_2$ is equal to $(k_1 + k_2 - 2)$ whereas in
$M_3$ that number is $k_1 + k_2 - 1$. Therefore, the moment $M_3$ is
smaller by a factor $1 / N$ than the product of the moments $M_1$ and
$M_2$. The difference is caused by the fact that a summation over the
energies~$E_\alpha$ is carried out separately and independently in
$M_1$ and in $M_2$ while in $M_3$ only a single such sum occurs. Each
of these sums is equivalent to a factor $N$. If the correlated
expression~(\ref{v1}) contains several correlated moments of the type
$M_3$, our argument applies to each one of them. We conclude that the
variance~(\ref{v1}) of the numerator of ${\cal F}(t)$ is small at
least of order $1 / N$ compared to the square of the average. We
observe that our argument is consistent with and generalizes the
result for the trace of the partition function in Section~\ref{par}.

Our argument makes no use of the specific form of the arguments of the
two traces in expression~(\ref{v1}). The argument holds for any two
traces that each contain the stochastic matrices $Y$. Therefore, it
applies likewise to any pair of traces in the numerator of
Eq.~(\ref{d5}). It follows that the variance of $C(t)$ is also small of
order $1 / N$ compared to $\langle C(t) \rangle^2$.

\section{Summary, Results, Discussion}

We have calculated the out-of-time ordered correlator $C(t)$ in
Eq.~(\ref{d5}) and the function ${\cal F}(t)$ in Eq.~(\ref{d6}) for a
chaotic many-body quantum system. Writing the Hamiltonian $H$ of the
system as the sum of a Hartree-Fock (HF) part plus a residual
interaction $V$, we have postulated that $V$ mixes the HF states
sufficiently strongly so that in the semiclassical regime, quantum
chaos is fully realized. We have quantified that assumption using the
BGS conjecture. We have postulated that for an energy interval
$\Delta$ located anywhere in the semiclassical regime, eigenvalues
$E_\alpha$ of $H$ and the associated eigenfunctions (defined by the
orthogonal matrices $O_{m \alpha}$) obey GOE statistics. We have
defined a random-matrix ensemble every member of which possesses these
statistical properties. Then both the out-of-time ordered correlator
$C(t)$ in Eq.~(\ref{d5}) and the function ${\cal F}(t)$ in
Eq.~(\ref{d6}) are stochastic processes. Using the simplifying
assumptions~(\ref{av2}) for the test operators $V$ and $W$, we have
worked out the ensemble averages and variances of both $C(t)$ and
${\cal F}(t)$. That is done using the assumption that the number of
states $N = \Delta \rho$ in the interval $\Delta$ obeys $N \gg
1$. Here $\rho$ is the average level density. We have calculated the
leading-order terms in an asymptotic expansion in inverse powers of
$N$.

We have shown that
\ba
\label{dis1}
{\cal F}(t) = \langle {\cal F}(t) \rangle \ , \ C(t) =
\langle C(t) \rangle
\ea
for almost all members of the random-matrix ensemble with the
exception of a set of measure zero. The measure is defined by the
integration measure for the random variables ${\cal E}_m$ and $O_{m
  \alpha}$. The statements on $\langle C(t) \rangle$ and on $\langle
{\cal F}(t) \rangle$ made below hold, therefore, for almost all
members of the random-matrix ensemble.

Results for $\langle C(t) \rangle$ and for $\langle {\cal F}(t)
\rangle$ are given in Eqs.~(\ref{av4}) and (\ref{av5}) and in the
Appendix.  The factor $\exp \{ \beta^2 \Delta^2 / 2 \}$ stemming from
$\langle {\rm Tr} (Z) \rangle$ in Eq.~(\ref{p1}) cancels completely in
$\langle C(t) \rangle$ while cancellation is incomplete in $\langle
{\cal F}(t) \rangle$. That is because $C(t)$ carries the factor $Z$
while ${\cal F}(t)$ carries four factors $Z^{1/4}$. Averaging yields
for $Z$ the factor $\exp \{ \beta^2 \Delta^2 / 2 \}$ and for $Z^{1/4}$
the factor $\exp \{ \beta^2 \Delta^2 / 32 \}$. Four such factors yield
$\exp \{ \beta^2 \Delta^2 / 8 \}$, short of cancellation. That shows
that ${\cal F}(t)$ is an auxiliary quantity only and not a natural
physical measure of quantum chaos.

We compare our results with those of Ref.~\cite{Mal16}. For large
times, $\langle {\cal F}(t) \rangle$ in Eq.~(\ref{av5}) tends
asymptotically towards zero. That is in line with Ref.~\cite{Mal16}
where it is stated that for large times, ${\cal F}(t)$ should become
“small”. For large $t$, $\langle C(t) \rangle$ tends towards the last
two terms in Eq.~(\ref{av4}). These contain, respectively, the thermal
average of $W^2$ (of $V^2$) to which the weighted trace of $V^2$ (of
$W^2$) is coupled by a Gaussian of width $\sqrt{2} \Delta$ that
carries the energy differences. In Ref.~\cite{Mal16} it is stated that
$C(t)$ should asymptotically be of order $2 \langle V V \rangle
\langle W W \rangle$ (the angular brackets here stand for the thermal
average). That can only be a semiquantitative estimate because every
term in $C(t)$ in Eq.~(\ref{d5}) carries a single factor $Z / {\rm Tr}
(Z)$ while $2 \langle V V \rangle \langle W W \rangle$ carries two
such factors. Moreover the two factors $\langle V V \rangle$ and
$\langle W W \rangle$ are independent and not coupled to each other as
in our case.
 
Our approach goes beyond Ref.~\cite{Mal16} in that it displays
explicitly the time dependence of $\langle C(t) \rangle$ and $\langle
{\cal F}(t) \rangle$. In both cases and aside from oscillatory terms,
the relevant factor is $\exp \{ - 2 t^2 \Delta^2 / \hbar^2 \}$, a
Gaussian of width $\hbar / (2 \Delta)$. (Here and in what follows we
abandon the convention $\hbar = 1$). The Gaussian form of that factor
follows directly via Fourier transformation from Eq.~(\ref{h4}). The
Gaussian form of $F({\cal E}_m - \overline{E}_\alpha)$ in
Eq.~(\ref{h4}), in turn, has been introduced in Ref.~\cite{Wei24} on
the basis of numerical evidence. An alternative possibility, valid
probably for weak chaos, is a Lorentzian with an exponential function
as Fourier transform. We mention that fact in order to indicate that
the Gaussian form of the energy-dependent factors in Eqs.~(\ref{av4})
and (\ref{av5}), although very likely, does not stand on absolutely
firm ground and probably does not hold for weak chaos.

The time scale for decay of $\langle {\cal F}(t) \rangle$ and of
$\langle C(t) \rangle$ is given by $\hbar / \Delta$. That statement is
independent of the form of the function $F({\cal E}_m -
\overline{E}_\alpha)$ in Eq.~(\ref{h4}) and follows via Fourier
transformation from the assumption~(\ref{h3}) that the argument of
$F({\cal E}_m - \overline{E}_\alpha)$ is the dimensionless ratio
$({\cal E}_m - \overline{E}_\alpha) / \Delta$. We consider that aspect
of the time dependence of $\langle {\cal F}(t) \rangle$ and of
$\langle C(t) \rangle$ to be valid quite universally. Indeed, it is
diffult to see how a function that confines $|{\cal E}_m -
\overline{E_\alpha}|$ to an interval of width $\Delta$ could be
written other than as a function of that ratio.

The parameter $\Delta$ must occur in every chaotic quantum system for
which the BGS conjecture holds. That suggests the existence in the
corresponding chaotic classical system of a parameter that is the
classical analogue to $\Delta$. The obvious candidate is the leading
Lyapunov coefficient $\lambda_{\rm max}$, a parameter of dimension
(time)$^{- 1}$. In Ref.~\cite{Arv91} the analysis of the spacing
distribution of $400$ levels of the quantum Sinai billard has led to a
conjecture on the connection between $\Delta$ and $\lambda_{\rm
  max}$. The conjecture is formulated in Eq.~(3) of
Ref.~\cite{Arv91}. In our notation, that equation reads
\ba
\label{dis2}
\Delta = \hbar \lambda_{\max} \ .
\ea
Eq.~(\ref{dis2}) connects for the Sinai billard the classical
parameter $\lambda_{\rm max}$ via Planck's constant with the purely
quantum parameter $\Delta$. We conjecture that Eq.~(\ref{dis2}) holds
in general for all chaotic systems to which the BGS conjecture
applies. For the chaotic many-body system defined in
Section~\ref{cha}, the linear relationship~(\ref{dis2}) is consistent
with the fact that an increase of $\Delta$ implies mixing of the HF
states over a larger energy interval and, thus, over a larger number
of HF states and, in that sense, stronger quantum chaos, while an
increase of $\lambda_{\rm max}$ implies stronger divergence of
classical trajectories in phase space and, thus, stronger classical
chaos.

Collecting our arguments we are led to the following generalized
formulation of the BGS conjecture:

{\it For time-reversal invariant dynamical systems whose classical
  analogues are fully chaotic, there exists for every value of the
  energy $E$ in the semiclassical regime an interval of width $\Delta$
  centered on $E$ within which the spectral fluctuation properties are
  the same as for the GOE. For energies below the semiclassical
  regime, the agreement is only approximate. The width $\Delta$ is
  related to the leading Ljapunov coefficient $\lambda_{\rm max}$ of
  the chaotic classical system by $\Delta = \hbar \lambda_{\rm max}$.}

The conjecture is based on data for the Sinai billard, as was the BGS
conjecture. Much of the work reviewed in Ref.~\cite{Gar23} was
motivated by the attempt to relate the time dependence of the
out-of-time ordered correlator for a chaotic quantum system to the
leading Ljapunov coefficient $\lambda_{\rm max}$ of the corresponding
classical system. Our conjecture establishes such a connection. It
implies that the time dependence of out-of-time ordered correlators
for chaotic quantum systems is, for large times, universally governed
by the dimensionless parameter $(\lambda_{\rm max} t)$.

The conjecture is not supported by the results of
Ref.~\cite{Gar18}. For a finite-size matrix model the authors show
that for large times, the decay of OTOC is exponential in time, with a
rate given by the Ruelle-Pollicot resonances. Apparently the authors
of Ref.~\cite{Gar18} did not analyse the spectral fluctuation
properties of the underlying quantum Hamiltonian. These form the basis
of the present approach, see Section~\ref{cha}. It is, therefore, not
clear whether the model of Ref.~\cite{Gar18} meets our assumptions
and, if so, where the discrepancy arises.

{\bf Acknowledgement.} The author is grateful to Eric Lutz for a
stimulating discussion, and for drawing his attention to
Ref.~\cite{Gar18}.

\section*{Appendix}

With $\chi = - \beta / 4 - i t$, the factor ${\cal F}_0$ in
Eq.~(\ref{av5}) is given by the sum of three terms. The first one is
\ba
\label{ap1}
&& {\rm Tr} \bigg( \exp \{ \chi^* H_{\rm HF} \} V \exp \{ \chi H_{\rm HF}
\} W \nonumber \\
&& \qquad \times \exp \{ \chi^* H_{\rm HF} \} V \exp
\{ \chi H_{\rm HF} \} W \bigg) \ .
\ea
The second term is
\ba
\label{ap2}
&& \sum_{m_1 m_3} \frac{1}{2 \sqrt{\pi} \rho({\cal E}_{m_1})} \exp \{
\chi ({\cal E}_{m_1} + {\cal E}_{m_3}) \} \\ && \times \exp
- ( {\cal E}_{m_1} - {\cal E}_{m_3})^2 / (4 \Delta^2) \} \langle W V
\rangle_{m_1 m_1} \langle W V \rangle_{m_3 m_3} \nonumber
\ea
where
\ba
\label{ap3}
\langle W V \rangle_{m m} = \sum_n W_{m n} \exp \{ \chi^* {\cal E}_n \}
V_{n m} \ .
\ea
The third term is obtained from the second one by complex conjugation
and the substitution $V \leftrightarrow W$.


\begin{thebibliography}{99}

\bibitem{Lar69}A. I. Larkin and Y. N. Ovchinnikov, JETP {\bf 28}, 1200
  (1969).

\bibitem{Gar23}I. Garcia-Mata, R. A. Jalabert, and D. A. Wisniacki,
  Scholarpedia {\bf 18(4)}, 55237 (2023).

\bibitem{Wei24}H. A. Weidenm{\"ul}ler, J. Math. Phys. A:
  Math. Theor. {\bf 57}, 165001 (2024).

\bibitem{Mal16}J. Maldacena, S. H. Shenker, and D. Stanford, JHEP
  2016, 106 (2016).

\bibitem{Boh84}O. Bohigas, M. J. Giannoni, and C. Schmit, Phys. Rev.
  Lett. {\bf 52}, 1 (1984). 

\bibitem{Meh04}M. L. Mehta, {\it Random Matrices} (3rd ed.), Academic
  Press, New York (2004).
  
\bibitem{Mon75}K. K. Mon and J. B. French, Ann. Phys. (N.Y.) {\bf 95},
  90 (1975).

\bibitem{Arv91}P. Arve, Phys. Rev. A {\bf 44}, 6920 (1991). 

\bibitem{Gar18}I. Garcia-Mata, M. Saraceno, R. A. Jalabert,
  A. J. Rongaglia, and D. A. Wisniacki, Phys. Rev. Lett. {\bf 121},
  210601 (2018).
  
\end{thebibliography}
\end{document}